\documentclass[preprint,showpacs,floatfix]{revtex4-1}
\usepackage[dvips]{graphicx}
\usepackage{epsfig}
\usepackage{color}
\usepackage[normalem]{ulem}
\usepackage{amsmath,amsfonts,amssymb,bm}
\setcitestyle{numbers,square}
\usepackage{verbatim}
\usepackage[capitalize]{cleveref}
\usepackage{color,soul}
\usepackage{sans}
\usepackage[labelformat=simple]{subfig}

\UseRawInputEncoding
\begin{document}
\title{Crucial effect of transverse vibrations on the transport through polymer chains}

\author{Alexei Boulatov} 
\affiliation{National Research University Higher School of Economics,  11 Pokrovsky Blvd.,  Moscow, Russia, 101000}

\author{Alexander L. Burin\thanks{Corresponding author}}
\email[]{email@email.com}
\affiliation{Department of Chemistry, Tulane University. New Orleans, LA 70118}

\date{\today}
\begin{abstract}
The low temperature transport  of electron, or vibrational or electronic exciton  towards polymer  chains turns out to be dramatically sensitive to its interaction with transverse acoustic vibrations. We show that this interaction leads to substantial polaron effect and decoherence, which are generally stronger than those associated with  longitudinal vibrations. For site-dependent interactions transverse phonons form subohmic bath leading to the quantum phase transition accompanied by full suppression of the transport at zero temperature   and fast decoherence characterized by temperature dependent rate $k_{2} \propto T^{3/4}$ at low temperature  while $k_{2} \propto T^{2}$ for site-independent interactions.   The latter dependence was used to interpret  recent measurements of temperature dependent vibrational energy transport in polyethylene glycol oligomers. 
\end{abstract}
\maketitle

% Changes references Monro16 and Smith2016  were to the same work so I put them together. Replaced reference 9 with that to Monroe where the modification of exponent has also been discussed. Added reference to Mossner17LR

\section{Introduction}

Coherent long-distant charge and energy transport in molecules is of a great fundamental and practical interest.
Electron transport is important in electron transfer reactions \cite{Marcus1,Marcus2} and for molecular electronics \cite{AVIRAMRatner1974277,NitzanScience03,Lu2009tunnelingtohopping,
Thomas2019ResonantMolJunc}.  Exciton  transport is significant in optoelectronic devices \cite{Excitonalvertis2020impact,ExcAcPhonon2017} and photosynthesis \cite{Maiuri2018PhotSynth}.  Vibrational energy relaxation and transport are generally responsible for the  chemical energy balance \cite{AbeScience07,15LeitnerReview,PandeyLeitner2016ThermSign} and they are of interest for modern molecular devices capable to transfer heat at nanoscales  \cite{Cui2019SingleMoleculeThermCond,Abe03,
15LeitnerReview,SegalReview2016}. 
Coherent long-distance transport can be naturally expected through periodic or quasi-periodic polymer chains; possible examples of such transport include electron and/or exciton transport in DNA \cite{ab04DNAReview,ab08ExcDNA,Conwell01PolDNA} and vibrational energy transport towards periodic oligomers \cite{Troe04BallFirstExp,DlottScience07,Abe03,SegalReview2016,Asegun08Polyethilenetransport, Rubtsov12Acc,ab15AccountsIgor,ab19IgorReview}. 

In periodic one-dimensional molecules  the coherent ballistic transport can be naturally expected because of the delocalization of particle states at sufficiently low temperature where the thermal fluctuations  are substantially frozen out. Indeed, it has been demonstrated even at room temperature that   the vibrational energy can propagate in a coherent manner towards  a long distances of tens of nanometers in various organic polymers  \cite{Troe04BallFirstExp,DlottScience07,
ab15ballistictranspexp,ab14PerFluoroAlkExp,
ab16jpcPEGs,ab20PegsExp} dissolved in weakly polar solvents. However, at room temperature coherence is usually broken for vibrational energy transport  in peptides dissolved in highly polar water \cite{HammBotan07peptides,YuLeitner03,Hamm09MD} and for electron or electronic exciton transport due to very high extent of decoherence  \cite{Maiuri2018PhotSynth} and/or  polar environment reorganization  \cite{ab08jcpDNALocaliz,ab12DNAChargeTdep}. Yet coherent transport of electrons should become significant at sufficiently low temperature of $10-100$K as demonstrated in recent measurements of resonant electron transport through molecular junctions \cite{Thomas2019ResonantMolJunc}.  The coherence is defined by the relationship of the effective transport bandwidth $\tilde{\Delta}$ and transport decoherence rate $k_{2}$. At low temperature both parameters are determined by  particle interactions with acoustic phonons by means of polaron formation reducing the bandwidth since the moving particle should be accompanied by the lattice disturbance around it, and inelastic interaction destroying coherence of transport. One-dimensional molecules possess the transverse phonon spectrum branch that is substantially different from the longitudinal one. Here we show that this branch can be dramatically significant for the particle transport.

At low temperatures, the decoherence usually takes place due to the interaction with acoustic vibrational modes (phonons) having the quantization energy $\hbar\omega$ of order of $k_{B}T$, which always exist due to their gapless spectrum . Acoustic vibrations of a linear chain split into longitudinal, torsional  and two transverse spectrum branches \cite{PolimVibr1994Book,
landau1986Elasticitytheory,NanoTubeVibr20071,
ExcAcPhonon2017}. To the best of our knowledge the consideration of vibrations has been substantially restricted to longitudinal phonons having sound wave spectrum characterized by a linear dispersion law. This dispersion law describes stretching (longitudinal) or torsional acoustic phonon spectrum branches. In the translationally invariant polymer molecule the interaction of particle with these modes lead to a transport decoherence due to two phonon processes with the rate decreasing with the temperature as $k_{T} \propto T^{3}$ \cite{KP86,GogolinPolaronReview88PhysRep,
KAGANProkofevReview1992}. For site dependent interactions there exist substantial polaron effect 
due to the infrared catastrophe similar to the transport of particle interacting with the ohmic bath  as it has been revealed using variational methods \cite{PolaronVarToyozawaOriginal1961,ToyozavaBook63,PolaronVarSubOhmic2010,
Emery74PolaronVar,polaronvar-silbey84,
PolaronVarZhao1997,Vojtaprl08SubohmTransCont,
PolaronVarSubOhmic2010,
Chin11PolarSC,Frenzel2013SubOhmProdSt,Bera14Subohm},   
analytical theory \cite{CaldeiraLeggett81,Guinea85PolaronRelax,LeggettPolaronRevModPhys,KP86,
Kagan_quantum_1992,kehrein96DiscSubOhm,Chin06DiscntScPolSubohm,Yu07Discont,
Chin11PolarSC}, renormalization group methods 
\cite{Vojtaprl08SubohmTransCont,weiss_quantum_2008,PVojtaSubOhm12ErrorsConsp} and quantum Monte-Carlo approach \cite{VojtaQuantMC09SubOhmBth}.

% two discontinuous and self cons variational to both, vojta monte carlo to numerics, vojta error to numerics 

Here we investigate the interaction of particle with transverse (bending)  acoustic vibrations. Using a simple two site approximation equivalent to the spin-boson problem valid for the nearest site hopping transport we demonstrate that for a site dependent interaction these vibrations act as a subohmic bath \cite{weiss_quantum_2008,LeggettPolaronRevModPhys,Vojtaprl08SubohmTransCont,
PolaronVarSubOhmic2010} with a spectral function power law exponent $s=1/2$.  This result  is consistent with earlier analysis of quasi-one dimensional nano-mechanical resonators \cite{Seoanez_2007TLSTransvVibr} and with spectral density measurements in various molecular systems \cite{Pachon14SubOhmicExp}.  

The subohmic nature of bath results in a substantial polaron effect for site dependent interactions. Particularly, a quantum phase transition should take place at zero temperature accompanied by a disappearance of polaron hopping at small bare hopping amplitude. Such transition has been earlier predicted for a spin-boson problem in the presence of interaction with subohmic bath \cite{LeggettPolaronRevModPhys,Vojtaprl08SubohmTransCont,AndersPRLsubohmic07,VillaresDynLoc07,
Alvermannprlsubohmic09,Nalbach10SubOhmicSpDyn,PolaronVarSubOhmic2010,
WilnerSubOhmic153,Magazz__2015SubOhmic} and one-dimensional molecular systems can be used to observe it experimentally and use it in molecular electronics, for instance, in molecular  switches \cite{Bissell1994MolSwitch}.

For a site independent interaction transverse phonons act as a superohmic bath with the spectral function power law exponent $s=3/2$ that does not lead to infrared catastrophe. Yet interaction with them results in a coherence breakdown. In that case the rate of decoherence is predicted to scale  with the temperature as $T^{2}$ in agreement with earlier analysis \cite{LeggettPolaronRevModPhys}, thus being more efficient compared to that for longitudinal phonons. This prediction is consistent with the decoherence rate temperature dependence  used to interpret recent measurements of temperature dependent energy transport in PEG oligomers \cite{ab20PegsExp} {where the propagation of a single optical phonon excited by the external infrared pulse and interacting with thermal molecular vibrations has been investigated.

The interactions of particle with high energy polar optical phonons with high frequency can result in a polaron formation  with higher reorganization energy compared to that for weakly polar acoustic phonons \cite{Holstein,Troici14PolaronHolst,
PolaronOptPhFormTh2018}.  
We consider low temperature regime where this effect is temperature independent because the optical phonons are substantially frozen out like in recent low-temperature measurements of resonant transport through molecular junctions \cite{Thomas2019ResonantMolJunc}. The interaction with optical phonons is supposed to be included into the definition of particle hopping amplitude while all temperature dependencies will be determined by the interaction with acoustic phonons. For electrons or electronic excitons this regime can be realized well below a room temperature but this is what we need to attain the purely coherent transport. 

The paper is organized as following. In Sec. \ref{sec:model} the  particle interaction with  transverse phonons is introduced and its subohmic or superohmic nature for site dependent or independent interaction is pointed out. In Sec. \ref{sec:decoh} the polaron effect originated from this interaction  is characterized using the simplified self-consistent model of Ref. \cite{Chin11PolarSC}. In. Sec. \ref{sec:Decoh1} the particle transport and decoherence are described considering its transition between adjacent unit cells mostly following the analysis of Ref. \cite{LeggettPolaronRevModPhys}. In. Sec. \ref{sec:exp} the results are compared with recent measurements of temperature dependent vibrational energy transport in PEG oligomers \cite{PandeyLeitner17ThermPEGs,ab20PegsExp}. The paper is resumed by a brief conclusion.

% redefine reorganization energy here add that for transport. Discuss alternating model. 

\section{Transverse Phonons}
\label{sec:model}

A linear molecule generally possesses four spectral branches of acoustic vibrations corresponding to four continuous symmetry transformations. They  include the longitudinal branch due to displacements along the chain, the torsional branch formed by rotations about the molecular axis $z$ and two transverse branches due to displacements in $x$ and $y$ directions ($u_{x}$ and $u_{y}$) perpendicular to the molecular axis \cite{PolimVibr1994Book,landau1986Elasticitytheory,
NanoTubeVibr20071,Nayak06nanotubes}. 

The interaction of longitudinal phonons  with the propagating particle in a pure one-dimensional system  has been considered  earlier in Refs.  \cite{KP86,KAGANProkofevReview1992} and interaction with torsional phonons can be treated similarly.  These interactions lead to the decoherence due to two phonon processes. This decoherence rate  behaves at low temperature as  $T^{3}$. Here we examine the interaction with the transverse branches. 

In contrast to longitudinal and torsional spectrum branches  transverse modes follow  a quadratic dispersion law expressing vibrational frequency $\omega_{q}$ in terms of the mode wavevector $q=2\pi k/(Na)$ as  $\omega_{q} = \omega_{0}(k/N)^2$, where $\omega_{0}$ represents a maximum bending vibrational frequency, integer number $k=-N/2,...N/2$ enumerates normal modes  and $a$ stands for a chain period. The quadratic dispersion law   is a consequence of the linear chain  geometry where the infinitesimal transverse displacement in $x$ direction linear with the coordinate $z$ along the molecular axis  is equivalent to the rotation about $y$ axis conserving a system energy.  For the sake of simplicity we use  periodic boundary conditions for chain vibrations and represent normal modes as  running waves $u_{kp}=e^{2\pi ik/N}/\sqrt{N}$. This assumption does not affect the results significantly compared to other boundary conditions at sufficiently large number of sites.

The transverse acoustic phonon  Hamiltonian can be expressed using raising and lowering operators for normal modes, enumerated by the normal mode index $k$ and branch index $\alpha=x$ or $y$ characterizing the mode polarization, as
\begin{eqnarray}
\widehat{H}_{l}=\sum_{k, \alpha=x,y} \hbar\omega_{k}\left(\widehat{b}_{k\alpha}^{\dagger}\widehat{b}_{k\alpha}+\frac{1}{2}\right), ~ \omega_{k}=\omega_{0}\left(\frac{k}{N}\right)^2.
\label{eq:LattNormModes}
\end{eqnarray}
We assume identical spectra and mutual independence for  modes with polarization in  $x$ and $y$ directions. An extension of theory  to a more general  case is straightforward. 

The true spectrum $\omega_{k}$ deviates from its definition in Eq. (\ref{eq:LattNormModes}) at large numbers $k$ approaching the number of atoms $N$. We will use a quadratic spectrum up to this maximum as in the Debye model \cite{Kittel2004}. Our estimates for temperature dependent decoherence rate is not sensitive to this assumption since the decoherence is due to long-wavelength vibrational modes (see Sec. \ref{sec:Decoh1}).  

The particle Hamiltonian can be written in a tight-binding approach, using its discrete state basis $|p>$ with integer index $p=1,...N$ enumerating unit cells, as 
\begin{eqnarray}
\widehat{H}_{p}=- \Delta\sum_{p=1}^{N-1}(\widehat{c}_{p}^{\dagger}\widehat{c}_{p+1}+\widehat{c}_{p+1}^{\dagger}\widehat{c}_{p}), 
%\nonumber\\
% n_{p}= \widehat{c}_{p}^{\dagger}\widehat{c}_{p}.
\label{eq:Part}
\end{eqnarray} 
where the operators $\widehat{c}_{p}, \widehat{c}_{p}^{\dagger}$ describe  annihilation and creation of the particle at site $p$, respectively,  and energy $\Delta$ stands for the particle hopping amplitude. The operators $\widehat{c}$ obey Fermi statistics for electrons and Bose statistics for vibrational or electronic excitons. Since we consider only one particle per the whole molecule, in agreement with the typical experimental conditions \cite{IgorPNASRA2DIR}, the results are not sensitive to that statistics. 

%Here and below we assume that the bandwidth for the particle transport is much than the thermal energy
%\begin{eqnarray}
%4\Delta < k_{B}T. 
%\label{eq:narrBand}
%\end{eqnarray}
%Otherwise the site basis is not applicable since the energetically favorable basis states should be represented by superposition of several site states. This regime needs a separate consideration that is beyond the scope of the present work.  

The interaction between the particle and long-wavelength transverse vibrations should  be expressed using the second derivatives of transverse displacement coordinates with respect to the molecular axis coordinate $z$ since  displacements linear with $z$ does not generate a stress being  equivalent to a rotation. For the sake of simplicity we assume that the particle  interacts with the only vibrations polarized along the $x$ axis. Then the second derivative of a displacement coordinate has the form $k^2 \sqrt{\hbar/(2M\omega_{k})}(b_{k}^{\dagger}+b_{-k})/N^{2+1/2}$ where $M$ is the mass of the unit cell \cite{Kittel2004}. 
Then using the definition of frequencies in Eq. (\ref{eq:LattNormModes}) one can express the particle-vibration interaction   as 
\begin{eqnarray}
\widehat{V}_{int}=-\frac{1}{\sqrt{2N}}\sum_{p, k} \eta_{p}\frac{k}{N}
\sqrt{\lambda \hbar\omega_{0}} n_{p}e^{i\pi kp/N}\left(\widehat{b}_{k}^{\dagger}+\widehat{b}_{-k}\right). 
\label{eq:part-phint}
\end{eqnarray}
where $n_{p}=c_{p}^{\dagger}c_{p}$ is a particle density operator in a site $p$, energy $\lambda$ is chosen to be the reorganization energy for the particle transition between adjacent sites as we will see below, while the dimensionless constants $\eta_{p}$ express interaction site-dependence. They  are chosen to be either $1$ or $-1$ to make all  sites $p$ equivalent. 

The choice $\eta_{p}=1$ corresponds to a site-independent interaction, while the alternating choice $\eta_{p}=(-1)^{p}$  corresponds to the 
site-dependent interactions that can take place for oligomers with a dihedral angle $\pi$. In a more general case of an angle $\varphi$ one can introduce the interaction to be proportional to the superposition of second derivatives of displacements in $x$ and $y$ directions with weights factors $\cos(p\varphi)$ and $\sin(p\varphi)$, respectively. Here we consider the cases of $\varphi=0$ or $\pi$, while the generalization to an arbitrarily angle $\varphi$ is quite straightforward.

Following Refs. \cite{LeggettPolaronRevModPhys,KP86} one can characterize the particle transport  considering  the particle transitions between two adjacent sites $p$ and $p+1$. Then the problem can be reformulated as a spin-boson problem in terms of Pauli matrices $\sigma^{x}$ and $\sigma^{z}$ with spin states $\sigma^{z}=\pm 1$ replacing the particle states $|p>$ and $p+1>$, respectively. Then the particle transport is described by the operator  $\Delta\sigma_{x}$  and the particle-phonon interaction, Eq. (\ref{eq:part-phint}),  takes the form 
\begin{eqnarray}
\widehat{V}_{int}=-\frac{\sigma^{z}}{2\sqrt{2N}}\sum_{p, k} \frac{k}{N}
\sqrt{\lambda \hbar\omega_{0}} (1\mp e^{2i\pi kp/N})\left(\widehat{b}_{k}^{\dagger}+\widehat{b}_{-k}\right), 
\label{eq:sp-ph}
\end{eqnarray}
where the sign minus corresponds to a site-dependent interaction, while the sign plus describes the site-dependent one.

The interaction of particle (spin) with the environment having the general form $\sigma^{z}\sum_{k}f_{k}(b_{k}^{\dagger}+b_{-k})$ can be characterized by its spectral density given by $J(\omega)=\sum _{k}|f_{k}|^2\delta(\omega-\omega_{k})$ \cite{LeggettPolaronRevModPhys,weiss_quantum_2008}. For the interaction in Eq. (\ref{eq:sp-ph}) this spectral density is given by
\begin{eqnarray}
J(\omega) =\frac{\lambda\hbar \omega_{0}}{8N}\sum_{k}\left(\frac{k}{N}\right)^2\mid 1\mp e^{i2\pi k/N}\mid^2\delta(\omega-\omega_{k}), 
\label{eq:SpectDens}
\end{eqnarray}
where the sum should be evaluated in a continuum spectrum limit of $N \rightarrow \infty$. 

The low frequency asymptotic behavior of $J(\omega)$ determines the low temperature particle transport. In the case of site-dependent interaction   one gets $J(\omega) \propto \omega^{1/2}$ that corresponds to the subohmic bath \cite{LeggettPolaronRevModPhys}. 
In the case of identical interactions the difference of two interaction constants acquires an additional factor $|1-e^{i2\pi k/N}|^2 \propto \omega_{k}$ and the spectral density shows the superohmic behavior  $J(\omega) \propto \omega^{3/2}$. In both cases the spectral density is larger then for longitudinal phonons (the ohmic bath, $J(\omega) \propto \omega$, for site-dependent interactions and $J(\omega) \propto \omega^2$ otherwise \cite{KP86}). It is noticeable that in two dimensions transverse out of plane phonons act as an ohmic bath that can lead to a substantial polaron effect. 

It is recognized \cite{LeggettPolaronRevModPhys} that the interaction with subohmic bath can lead to the quantum phase transition and both subohmic and superohmic baths with spectral density exponents $s=1/2$ or $3/2$ breaks down the coherence at a finite temperature. Below we examine the phase transition, and  decoherence and transport rates in various regimes of interest. 

The only diagonal terms with respect to the particle site population ($\propto n_{p}$)  are considered for the interaction  with vibrations following the standard approach \cite{Holstein,BixonJortner,KP86}. We treat  the PEG oligomers considered in Sec. \ref{sec:exp}  using site-independent interactions. 

We ignore the influence of molecular environment on molecular vibrations and particle dynamics. This assumption is justified in molecular junctions \cite{Abe03,Asegun08Polyethilenetransport,
Thomas2019ResonantMolJunc}, if different molecules are located far enough from each other, or in the gas phase experiments  \cite{GasPhVibrreview15,GasPhaseVbrBrnken2019,GasPhaseVibrBakels2020}.  
It can also be reasonably justified for the interaction with liquid or solid solvent because intramolecular interactions substantially exceed  intermolecular ones and because of extremely slow decoherence due to three dimensional phonons at low temperature. 
Indeed,  a decoherence rate due to interaction with three dimensional phonons decreases with the temperature  as $T^{7}$  \cite{KaganMaksimov73}.  The rates associated with the molecular vibrations decrease much slower ($T^{2}$ or $T^{3}$ \cite{KP86}). Environment was not significant in the seminal work by Stewart and McDonald \cite{Stewart} where the molecules were dissolved in a liquid helium. 

For solid solvent and very low frequency modes the consideration should fail because of the molecular motion constraints imposed by the host solid. We hope that this limit is not reached in the experiments \cite{ab20PegsExp} discussed below in Sec. \ref{sec:exp}; yet this assumption is the subject for the future verification.

\section{Polaron effect}
\label{sec:decoh}

% Introductory words

The particle placed into a unit cell disturbs equilibrium positions of nearby atoms by means of the interaction in Eq. (\ref{eq:part-phint}). When it hops to the neighboring chain site the equilibrium positions  of atoms change, which leads to the dynamic interaction of the particle with vibrations. 

At zero temperature  the transport of particle can be described as a polaron transport, meaning that the particle propagates together with the lattice disturbance. The polaron hopping between adjacent sites can be approximately characterized by the product of the particle hopping amplitude $\Delta$ and the overlap integral of reorganized lattice ground states %$|p>_{l}$ and $|p+1>_{l}$
 for particle being in sites $p$ and $p+1$  \cite{CaldeiraLeggett81,KAGANProkofevReview1992}. 
First, we characterize the atomic displacements in Sec. \ref{sec:displacements} and then evaluate their 
 overlap integral in Sec. \ref{sec:polaron}. 

At a finite temperature the small fraction of excited vibrations with energies of order of a thermal energy $k_{B}T$  can be scattered inelastically by the moving particle. The interaction of particle with these vibrations can lead to the decoherence of particle dynamics. We investigate this decoherence  considering a particle transition between adjacent sites in Sec. \ref{sec:Decoh1} using the generalized Fermi Golden rule \cite{Holstein,BixonJortner}. 

\subsection{Atomic displacements}
\label{sec:displacements}

% explain what are equilibria, reexpress potential energy in terms of coordinates and find minimum

To characterize the equilibrium state of lattice for the particle localized at a chain site $p$ we can temporarily set the particle hopping amplitude $\Delta$ in Eq. (\ref{eq:Part}) to zero and consider the particle at site $p$ suggesting the particle density to be $n_{i}=1$ for $i=p$ and zero otherwise.  Then the potential energy associated with the specific normal coordinate $u_{q}$  can be expressed as  %shown in Eq. (\ref{eq:part-phint}) 
(we skip polarization index $x$ since all further consideration is related to this polarization only)
\begin{eqnarray}
\frac{M\omega_{k}^2 u_{k}u_{-k}}{2} -\eta_{p}\sqrt{\lambda M\omega_{0}^2}\left[\frac{k^2}{N^2\sqrt{N}} u_{k}e^{2i\pi kp/N}+H.C. \right]. 
\label{eq:eqildisp1}
\end{eqnarray}
The equilibrium displacements of coordinates  $\delta u_{kp}$ for specific mode $k$ due to the presence particle at site $p$ can be found by minimizing the energy in Eq. (\ref{eq:eqildisp1}). Then we get %in the long-wavelength limit $k\rightarrow 0$ 
\begin{eqnarray}
\delta u_{kp}  \approx \eta_{p} \sqrt{\frac{\lambda}{M\omega_{0}^2}}\frac{e^{2i\pi kp/N}N^2}{\sqrt{N}k^2}. 
\label{eq:Displ}
\end{eqnarray}

%The new vibrational energy minimum is smaller than that in the absence of particle  by the reorganization energy $\lambda_{p}$ that can be expressed as 
%\begin{eqnarray}
%\lambda_{p}=\sum_{k} \frac{M\omega_{k}^2 |\delta u_{kp}|^2}{2}  \approx \frac{\gamma_{p}^2}{2M\omega_{0}^2a^2}.
%\label{eq:reorgEn}
%\end{eqnarray}
%This energy estimates the energy gain due to the local lattice rearrangement. The polaron effect is significant when this energy  exceeds the gain in energy $\Delta$ due to the pure hybridization of two particle states. 

The displacement of the specific atom at the position $x$ induced by the particle placed into the site $p$ ($u_{p}(x)$)  can be expressed evaluating the inverse Fourier transform of Eq. (\ref{eq:Displ}) in the large distance limit $|x-pa|\gg a$ as 
\begin{eqnarray}
\delta u_{p}(x) =\frac{a\eta_{p}}{2\pi \sqrt{N}}\sum_{k}\delta u_{kp}e^{-2i\pi kx/(Na)} 
%\approx u(p) + \frac{\gamma_{p}}{2\pi M\omega_{0}^2a^2}\int_{-\infty}^{\infty} \frac{(1-cos(q(x-p)))dq}{q^2}
%\nonumber\\
=u(p)+\sqrt{\frac{\lambda}{M\omega_{0}^2}}|x-pa|, 
\label{eq:Displ1}
\end{eqnarray}
where we used the identity $\pi=\int_{-\infty}^{\infty}dx (1-\cos(x))/x^2$  \cite{GradshteynRyzhik07}  and the integration over $k$ is extendable to infinite limits for $|x-p| \gg 1$.

The displacement at the site $p$ where the particle is located can be set to zero. Then the shape of the molecule affected by the particle becomes angular with the vertex located at the position of the particle as shown schematically by double lines in Fig. \ref{fig:DMRes} for sites $p$ and $p+1$.  The angle $\phi$ between the original molecular chain axis and its new direction is defined as $\tan(\phi)=\sqrt{\frac{\lambda}{M\omega_{0}^2}}/2$. 

The consideration ignores horizontal displacements that is justified for small angles $\phi$ only. This restricts our consideration to small interactions $\lambda \ll M\omega_{0}^2 a^2$. Since the typical value of the right hand side of this inequality is around $10$eV its violation for the reorganization energy $\lambda$ will lead to full suppression of transport due to the huge polaron exponent $e^{-\lambda/(\hbar\omega_{0})}$ because the quantization energy $\hbar\omega_{0}$ should be around $0.01$eV.

\begin{figure}
\includegraphics[scale=0.5]{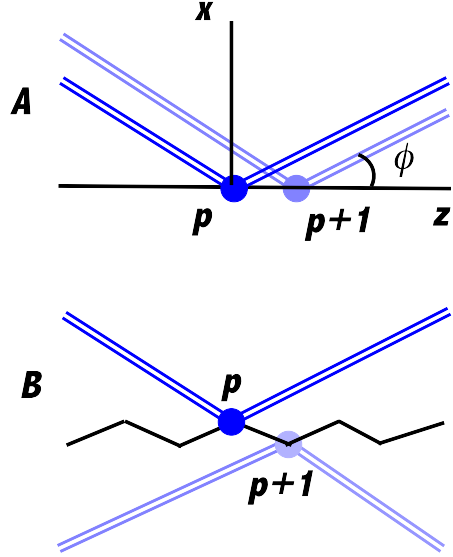} 
\caption{Reorganization of the chain shape due to the interactions with the particle for site-independent (top chart A) and site-dependent  (bottom chart B) interactions.  Double  lines (blue color lines online) schematically show bent chain axes due to the particle presence in sites $p$ or $p+1$. Vertical and horizontal $y$ and $z$ axes are indicated in  the top chart.}
\label{fig:DMRes}
\end{figure}

The particle transition between adjacent sites should lead to a simultaneous change in the molecular shape as shown in Fig. \ref{fig:DMRes} with change in the angle in the case of different interactions at adjacent  sites ( the case of alternating interactions in Fig. \ref{fig:DMRes}.B). These changes result in the polaron effect and decoherence as described below.

\subsection{Transverse Polaron}
\label{sec:polaron}

For the sake of simplicity we consider the particle transfer between two adjacent sites $p$ and $p+1$ which can be reduced to the spin-boson problem, Eq. (\ref{eq:sp-ph}). Following a standard procedure  one can apply the transformation   $\widehat{H}\rightarrow e^{i\widehat{S}}\widehat{H}e^{-i\widehat{S}}$ to the spin-boson Hamiltonian  with the operator $\widehat{S}$ defined as 
\begin{eqnarray}
\widehat{S}=\sigma^{z}\widehat{s}, ~ \widehat{s}
=\frac{1}{2\sqrt{2N}}\sum_{k}\sqrt{\lambda\hbar\omega_{k}}e^{i2\pi kp/N}(1\mp e^{i2\pi k/N})
\frac{\widehat{b}_{k}^{\dagger}-\widehat{b}_{-k}}{i\hbar\omega_{k}}.
\label{eq:CanTr1}
\end{eqnarray}
Remember that the sign minus corresponds to the site independent  interactions, while the sign plus to the different one. 

The transformation removes interaction of particle with vibrations, while  it modifies the hopping term as
\begin{eqnarray}
\widehat{V}_{h}=\Delta \left(\sigma^{+}e^{i2\widehat{s}}+H. C.\right). 
\label{eq:CanTr2}
\end{eqnarray}
and subtracts a quarter of  reorganization energy  $\lambda$ for the particle transition. It is straightforward to verify that this energy coincides with our definition of $\lambda$ in Eq. (\ref{eq:sp-ph}).   

The exponential operator in the $p \rightarrow p+1$ hopping term eliminates the equilibrium lattice displacement corresponding to the particle in site $p$ and creates the equilibrium displacement for the site $p+1$. It affects the hopping amplitude. For instance the average hopping amplitude can be introduced  as  $<\Delta e^{i2\widehat{s}}>=\Delta e^{-2<s^2>}$  where averaging is performed over non-interacting vibrations. However, for ohmic or subohmic baths this average in exponent diverges in the limit of infinitely long chain \cite{CaldeiraLeggett81,KP86}. This divergence can be overcome by removing the contribution of very slow vibrations with frequencies less than the coherent oscillation frequency of particle between two sites which can be estimated in terms of the renormalized hopping amplitude $\tilde{\Delta}$ as $\tilde{\Delta}/\hbar$. Consequently, they cannot follow particle transitions. 

The self-consistent analytical calculation of the renormalized hopping amplitude at zero temperature has been developed  using the advanced variational approach in Refs. \cite{Chin06DiscntScPolSubohm,Yu07Discont,Chin11PolarSC}. Following Ref. \cite{Chin11PolarSC} we can express the renormalized hopping amplitude $\tilde{\Delta}$ using their simplified self-consistent equation 
\begin{eqnarray}
\tilde{\Delta} =\Delta \times \exp\left[-\frac{\lambda\hbar\omega_{0}}{4N}\sum_{k}  \left(\frac{k}{N}\right)^2\frac{ |1\mp e^{2i\pi k/N}|^2}{(\hbar\omega_{k}+2\tilde{\Delta})^2}\right].
\label{eq:SCPolaron}
\end{eqnarray}
The additional factor of $2$ multiplied by the coupling strength $\tilde{\Delta}$ in the right hand side of Eq. (\ref{eq:SCPolaron}) compared to Ref. \cite{Chin11PolarSC} is due to the difference in the definitions of hopping terms.

Eq. (\ref{eq:SCPolaron}) is employed below to the analysis of the effective hopping amplitude $\tilde{\Delta}$ including the quantum phase transition for site-dependent interactions accompanied by vanishing this amplitude in the infinite system limit and zero temperature. One should notice that we skipped the symmetry breaking parameter between two spin states $\sigma^{z}=\pm 1$ introduced in Ref. \cite{Chin11PolarSC}. Since we consider a finite system a symmetry breaking can be problematic there. This parameter can be also responsible for the effect of higher energy states skipped in the variational approach used to derive Eq. (\ref{eq:SCPolaron}). Based on the previous analytical and numerical studies \cite{Chin06DiscntScPolSubohm,Chin11PolarSC,VojtaQuantMC09SubOhmBth}  we believe that the omitted effects are important only in the very narrow transition domain, while our approach should work everywhere else, although there is a substantial controversy about the relevance of various approaches \cite{PVojtaSubOhm12ErrorsConsp}.

Eq. (\ref{eq:SCPolaron})  should is  treated below separately for site  independent (Sec. \ref{sec:pol-ident}) and  size-dependent   (Sec. \ref{sec:pol-diff}) interactions with transverse vibrations. 

\subsubsection{Site independent interactions.}
\label{sec:pol-ident}

%Consider a zero temperature limit where a hypercotangent in Eq. (\ref{eq:selfcons}) can be set to $1$.  {\color{red} In that case the minimum  frequency constraint is determined by the inverse renormalized tunneliung amplitude as stated by Eq. (\ref{eq:zeroTFreqMin}). Consequently, two distinguishable regimes of maximum frequency $\omega_{0}$ smaller or larger than $\tilde{\Delta}/\hbar$ are discussed below. }

For a truly periodic chain with site-independent  interactions the sum in the exponent of  Eq. (\ref{eq:SCPolaron}) converges in the low frequency  limit $k\rightarrow 0$. Consequently, no infrared catastrophe takes place similarly to the case of longitudinal vibrations with identical site interactions \cite{KP86}. Then the finite renormalization of particle hopping amplitude  is expected. 

The renormalization can be significant or not depending on system parameters. Ignoring the lower frequency cutoff ($\tilde{\Delta}$) in the right hand side of Eq. (\ref{eq:SCPolaron})  one can estimate the minimum possible effective hopping amplitude  evaluating numerically the integral in exponent in the limit $N\rightarrow \infty$ as
$\lambda/(\hbar\omega)\int_{-1/2}^{1/2}dx \sin(\pi x)^2/x^2  \approx 0.95\frac{\lambda}{\hbar\omega_{0}}$. This yields  $\tilde{\Delta}_{\rm min} =\Delta e^{-\frac{0.95\lambda}{\hbar\omega_{0}}}.$.

%$<\widehat{s}^2>=\lambda/(\hbar\omega_{0})$. 

Then two distinguishable regimes are possible including large effective tunneling amplitude $\tilde{\Delta}>\hbar\omega_{0}$ where renormalization is negligible and one has $\tilde{\Delta} \approx \Delta$,  while in the opposite case one has 
\begin{eqnarray}
\tilde{\Delta} =\Delta e^{-\frac{0.95\lambda}{\hbar\omega_{0}}}.
\label{eq:hoampren}
\end{eqnarray}

Further analysis depends on the relationship between the reorganization energy $\lambda$ and the maximum  quantization energy $\hbar\omega_{0}$. For relatively small reorganization energy 
\begin{eqnarray}
\lambda<\hbar\omega_{0}
\label{eq:smalllambda}
\end{eqnarray}
a renormalization always gives a minor correction to a hopping amplitude. 

In the opposite case 
\begin{eqnarray}
\hbar\omega_{0} < \lambda, 
\label{eq:largelambda1}
\end{eqnarray}
a renormalization, Eq. (\ref{eq:hoampren}), definitely takes place for $\Delta < \hbar\omega_{0}$, while the situation is more complicated in the intermediate  case of $\hbar\omega_{0}<\Delta < \hbar\omega_{0}e^{0.95\lambda/(\hbar\omega_{0})}$. In this case a self-consistent problem in Eq. (\ref{eq:SCPolaron}) can have two solutions given by either Eq. (\ref{eq:hoampren}) or just by $\tilde{\Delta} \approx \Delta$. 

The right solution can be chosen minimizing the ground state energy  \cite{Nengji15AdiabPol,ab08jcpDNALocaliz}. The lack of renormalization gives a gain in energy of order of the bandwidth $\Delta$ and loss of order of the reorganization energy $\lambda$. It turns out that the renormalization of hopping amplitude  will be significant  for $\Delta < \lambda/2$ as can be demonstrated using the energy minimization within adiabatic \cite{Nengji15AdiabPol,ab08jcpDNALocaliz} or semiclassical  approaches \cite{StockSemiclassPol}. 

The behavior  of effective hopping amplitude $\tilde{\Delta}$ is summarized in Table \ref{tab:renorm-ident} and in Fig. \ref{fig:TunSup}, where the numerical solutions of Eq. (\ref{eq:SCPolaron})  are shown for the renormalized coupling strength $\tilde{\Delta}$ for different reorganization energies.  These solutions are consistent with our expectations. It is noticeable that there can be two solutions for large reorganization energies $\lambda\geq \hbar\omega_{0}$ and $\Delta \approx \lambda/2$ and the one possessing the minimum energy has been selected. This bi-stable behavior and nearly discontinuous transition  in Figs. \ref{fig:TunSup} c, d can be an artifact of limited relevance of Eq. (\ref{eq:SCPolaron}). However, we can definitely expect an abrupt change in the effective coupling strength in the vicinity of the point $\Delta=\lambda/2$, that can be of interest for practical applications like molecular switches \cite{Bissell1994MolSwitch}. 

\begin{table}
 \caption{Renormalization of hopping amplitude for identical site interactions at zero temperature limit.} 
 \label{tab:renorm-ident}
%  \begin{tabular}{|p{3.5cm}|p{3.5cm}|p{3.5cm}|}
\centering
\begin{tabular}{|l|c|c|c|r|}
  \hline
 $\lambda$ vs. $\hbar\omega_{0}$ &\multicolumn{2}{|c|}{$\lambda<\hbar\omega_{0}$}&\multicolumn{2}{|c|}{$\hbar\omega_{0}<\lambda$}\\
  \hline
$\Delta$   &$\Delta<\hbar\omega_{0}$ & $\hbar\omega_{0}<\Delta$ & $\Delta<\lambda/2$ & $\lambda/2 < \Delta$  \\
   \hline
$\tilde{\Delta}$ &  $\Delta e^{-\frac{0.95\lambda}{\hbar\omega_{0}}}$ & $\Delta$ & $\Delta e^{-\frac{0.95\lambda}{\hbar\omega_{0}}}$ & $\Delta$  \\ 
   \hline  
  \end{tabular}
\end{table}

%Since the spectrum is not specified at large wavevectors $q$ one can only estimate the exponent $<\widehat{s}^2>$ evaluating the integral  in Eq. (\ref{eq:zeroTpolaron}) using a small $q$ expressions as $<\widehat{s}^2> \approx \lambda/(\hbar\omega_{0})$. Consequently, one can expect the hopping amplitude renormalization as 

%This renormalization is determined by high frequency vibrations with $\omega_{q} \sim \omega_{0}$. The polaron integral in Eq. (\ref{eq:zeroTpolaron}) can include only vibrations with frequencies exceeding the transition rate between two sites \cite{KP86} that can be estimated as $\tilde{\Delta}/\hbar$. Then the inequality $\tilde{\Delta}/\hbar < \omega_{0}$ must be satisfied to allow the polaron formation. This inequality is definitely valid for sufficiently hopping amplitudes satisfying  Eq.  (\ref{eq:IneqPolForm}). This is the regime of interest of the present paper. Otherwise,  delocalization of a particle through many unit cells is significant.  

\begin{figure}[h]
\centering
\subfloat[]{\includegraphics[scale=0.4]{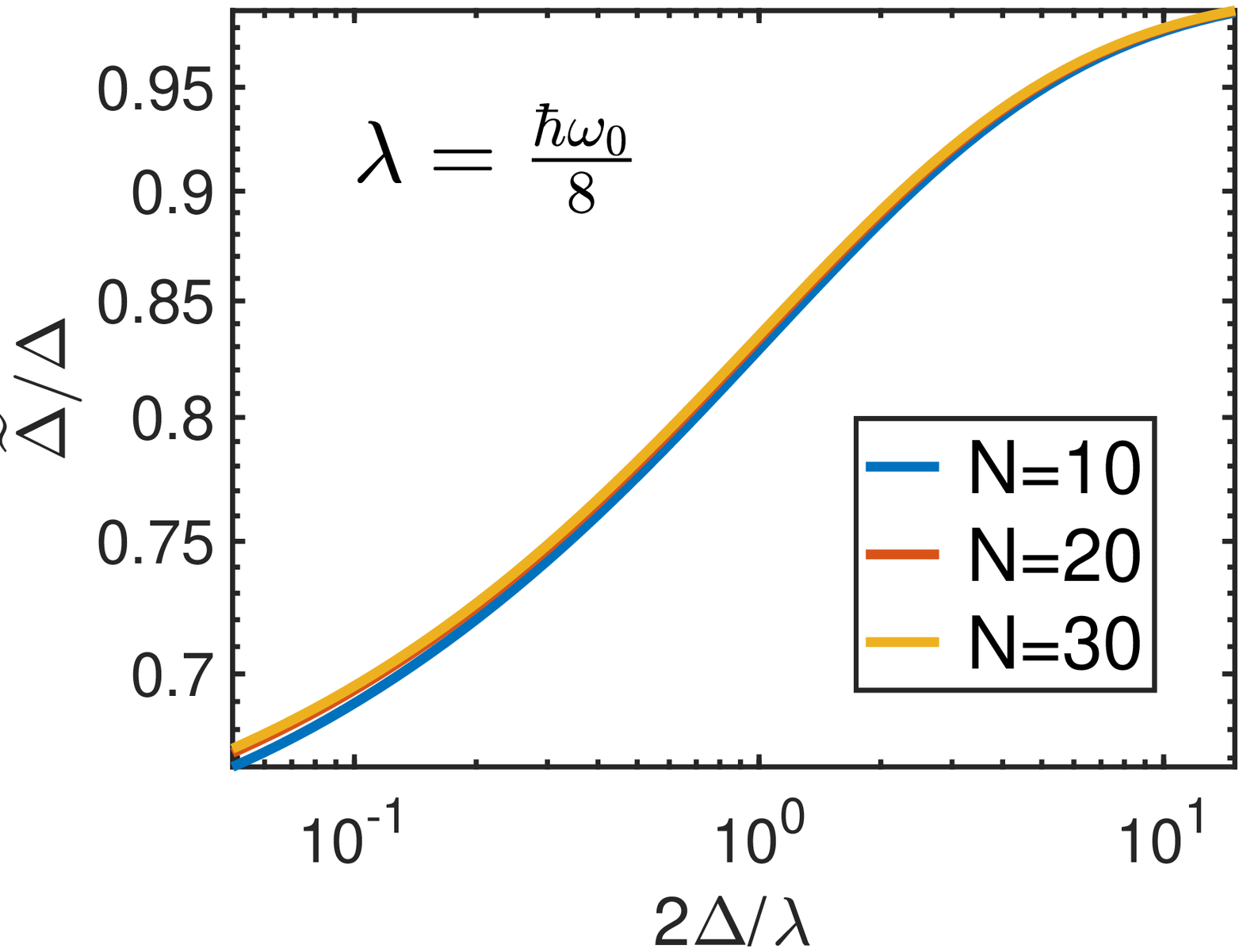}}
\subfloat[]{\includegraphics[scale=0.4]{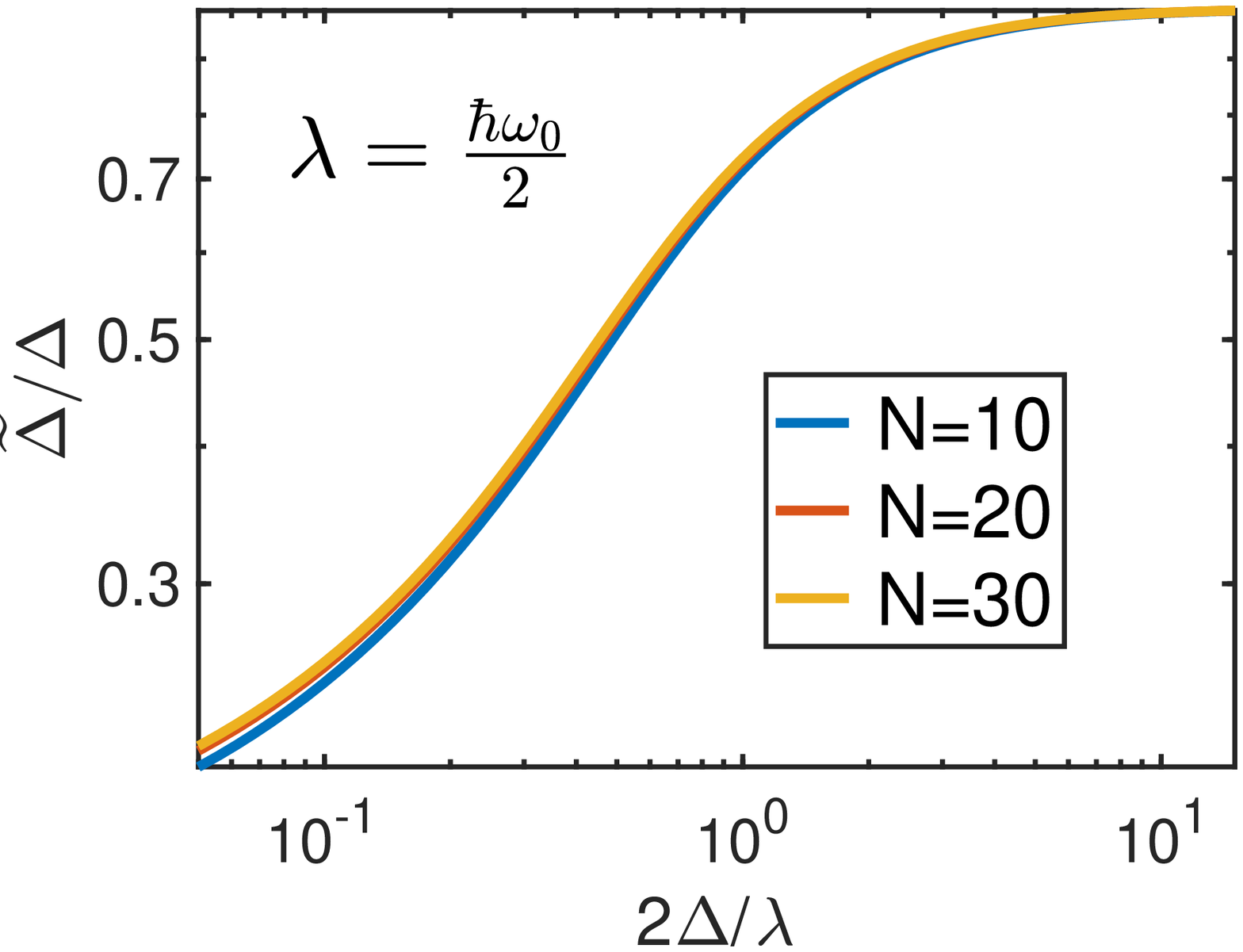}}
\\
\subfloat[]{\includegraphics[scale=0.4]{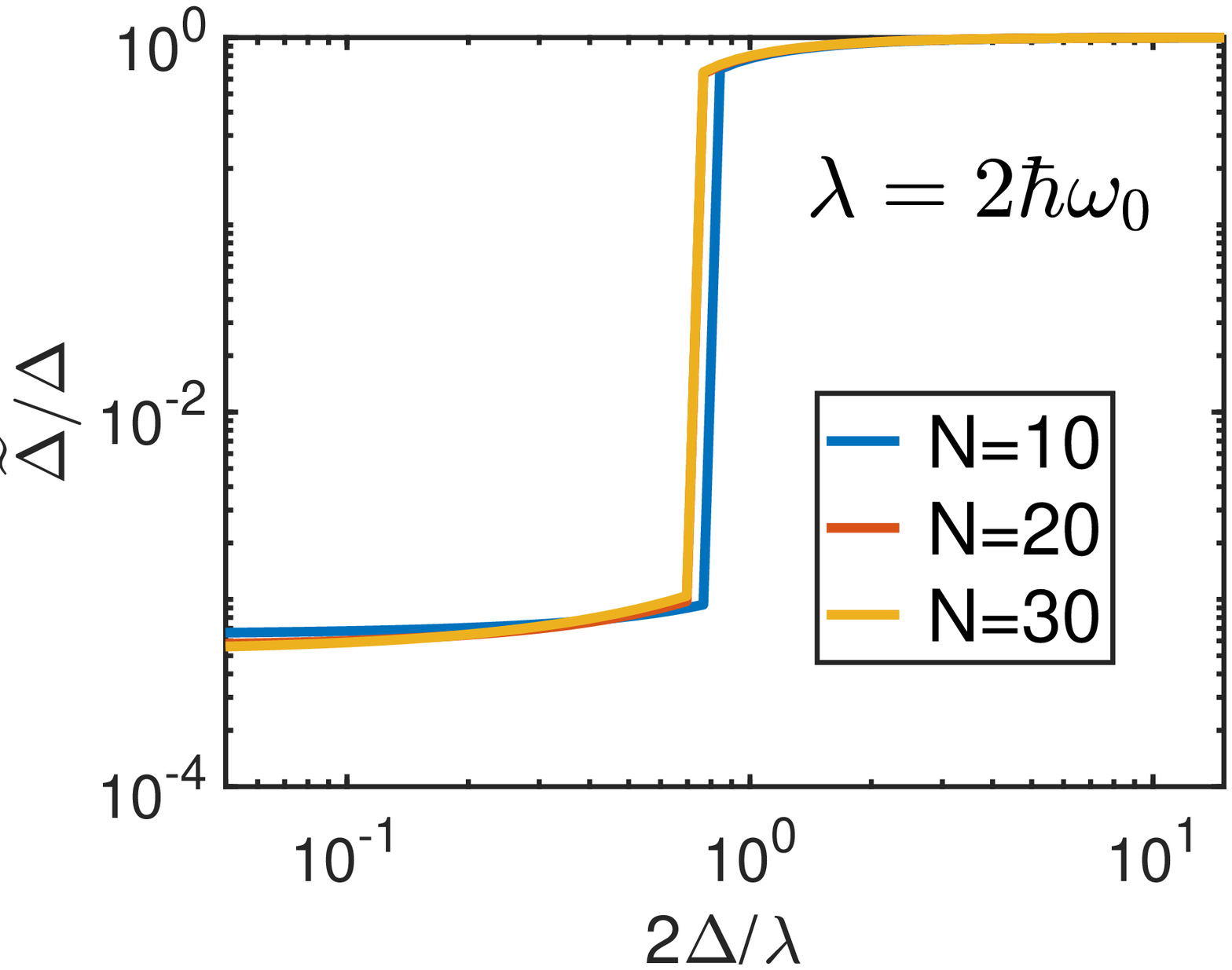}}
\subfloat[]{\includegraphics[scale=0.4]{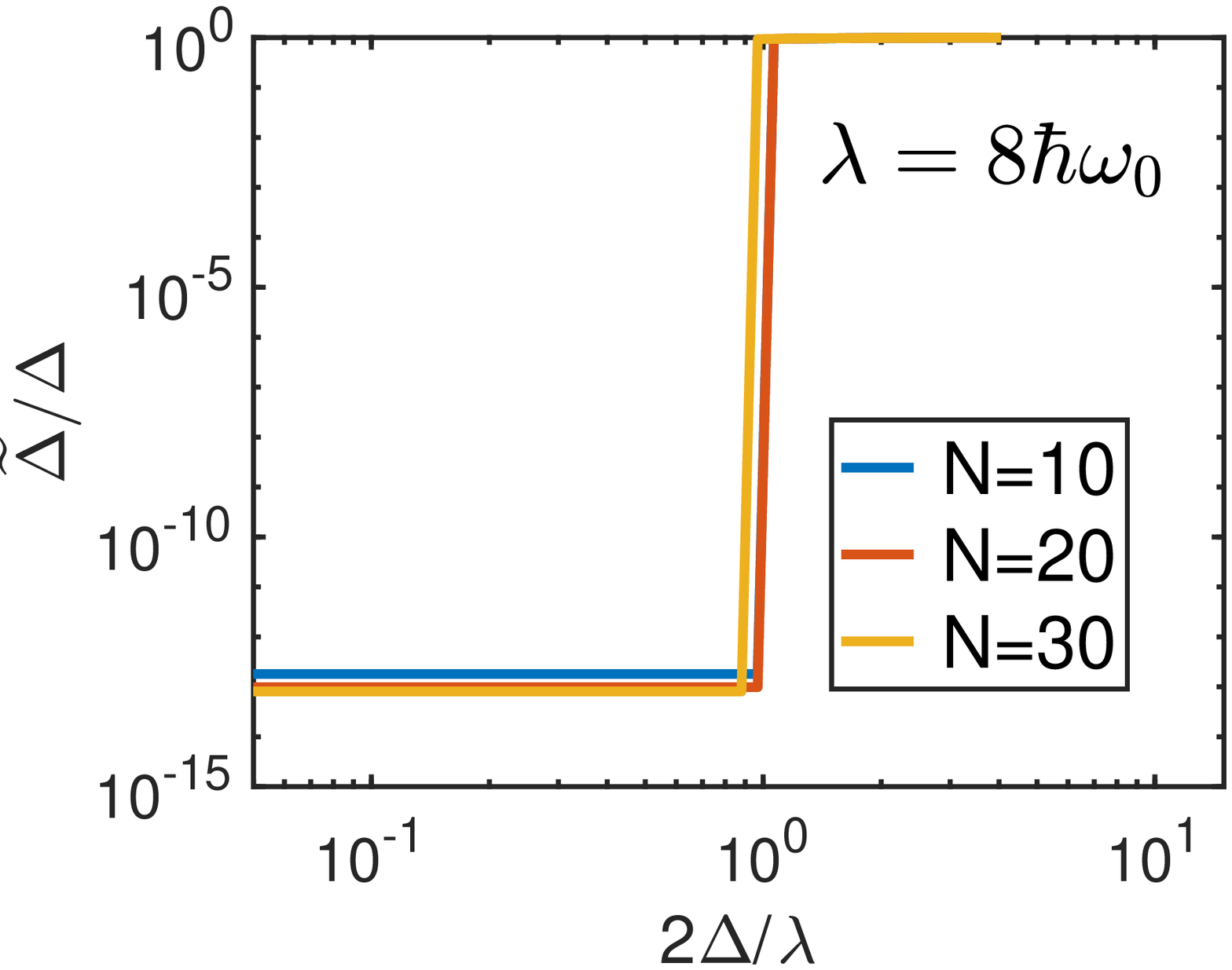}}
{\caption{\small  Relative renormalized hopping amplitude  $\tilde{\Delta}/\Delta$ vs. the relative bare hopping amplitude ($2\Delta/\lambda$) for different reorganization energies $\lambda/(\hbar\omega_{0})=1/8, 1/2, 2, 8$ shown in graphs {\bf a-d.} respectively, in the case of site independent interactions (color online, curve's height increases with increasing $N$)..} 
\label{fig:TunSup}}
\end{figure}

%Therefore at low temperatures the straightforward replacement of $\Delta$ with $\tilde{\Delta}$ will account for this renormalization. The parameter $I$ can vary from system to system \cite{KAGANProkofevReview1992}. We assume that it is less or equal unity; in the case of $I \gg 1$ the transport is fully suppressed at low temperature. 

At a finite temperature the  self-consistent equation, Eq. (\ref{eq:SCPolaron}), can be modified adding the phonon population factor $\coth(\hbar\omega_{k}/(2k_{B}T))$ to the expression under the sum sign. 
This modified sum diverges in the low frequency limit $k\rightarrow 0$ indicating the breakdown of coherence due to thermal phonons in contrast to that for longitudinal vibrations \cite{KP86} where two-phonon processes are needed for decoherence. Consequently, at a finite temperature, interaction of a particle with transverse vibrations leads to decoherence of the particle  transport. We will investigate this decoherence in Sec. \ref{sec:Decoh1}.

Below we discuss the case of site-dependent interaction at different sites, where there is  the infrared divergence of the sum in exponent in Eq. (\ref{eq:SCPolaron})  even at zero temperature that results in a quantum phase transition \cite{LeggettPolaronRevModPhys}.

\subsubsection{
 Site dependent interactions.}
\label{sec:pol-diff}

% refer to the graph show what exactly is going on. 
If particle interaction with transverse vibrations is different in adjacent sites then the expression for  the polaron exponent, in Eq. (\ref{eq:SCPolaron}) diverges at low frequencies  ($k\rightarrow 0$) even at zero temperature. The divergence is stronger compared to that for longitudinal phonons reflecting the subohmic nature of the phonon bath characterized by the spectral density exponent $s=1/2$, see Eq. (\ref{eq:SpectDens}). The lattice state overlap integral vanishes exponentially with the system size in contrast to its power law decrease for site-dependent longitudinal interactions. 

This is the consequence of dramatically different chain reorganizations at adjacent positions of the particle for different coupling constants.  As shown in Fig. \ref{fig:DMRes}.B  reorganization of atoms located infinitely far from the particle requires their infinite displacements in contrast with the case of identical interaction constants shown in Fig. \ref{fig:DMRes}.A. 

In the infinite system limit $N\rightarrow \infty$ Eq. (\ref{eq:SCPolaron}) predicts vanishing of renormalized hopping amplitude at small bare amplitudes $\Delta$ in a full accord with earlier considerations \cite{LeggettPolaronRevModPhys,VojtaQuantMC09SubOhmBth,
Chin06DiscntScPolSubohm,Chin11PolarSC} which suggest the quantum phase transition at some critical coupling strength $\Delta=\Delta_{c}$. This means $\tilde{\Delta}=0$ at $\Delta<\Delta_{c}$ and it is finite otherwise.

The transition point is sensitive to the relationship between the reorganization energy $\lambda$ and the maximum quantization energy $\hbar\omega_{0}$. To the best of our knowledge most studies  were dedicated to the case of small reorganization energy $\lambda \ll \hbar\omega_{0}$. In this limit the analysis of Eq. (\ref{eq:SCPolaron}) leads to the  estimate of the critical coupling strength as \cite{Chin06DiscntScPolSubohm} 
\begin{eqnarray}
\Delta_{c} \approx \frac{\pi^2e^{2}}{128}\frac{\lambda^2}{\hbar\omega_{0}}. 
\label{eq:Delta_c1}
\end{eqnarray}
This result can be obtained evaluating the integral in exponent of the self-consistent equation, Eq. (\ref{eq:SCPolaron}) assuming that the integral is determined by small frequencies that permits us to set $1+e^{2\pi ik/N}\approx 2$ and extend the upper integration limit to infinity. Then it can be evaluated as $f(\tilde{\Delta})=2\sqrt{\Delta_{c}/\tilde{\Delta}}/e$. Then Eq. (\ref{eq:SCPolaron}) takes the form $\Delta/\Delta_{c}=4e^{-2}e^{\sqrt{f}}/f$. The right hand side of the latter equation has a minimum equal unity that estimates the critical coupling for the quantum phase transition because for $\Delta/\Delta_{c}<1$ the equation has no solutions. 

\begin{figure}[h]
\centering
\subfloat[]{\includegraphics[scale=0.4]{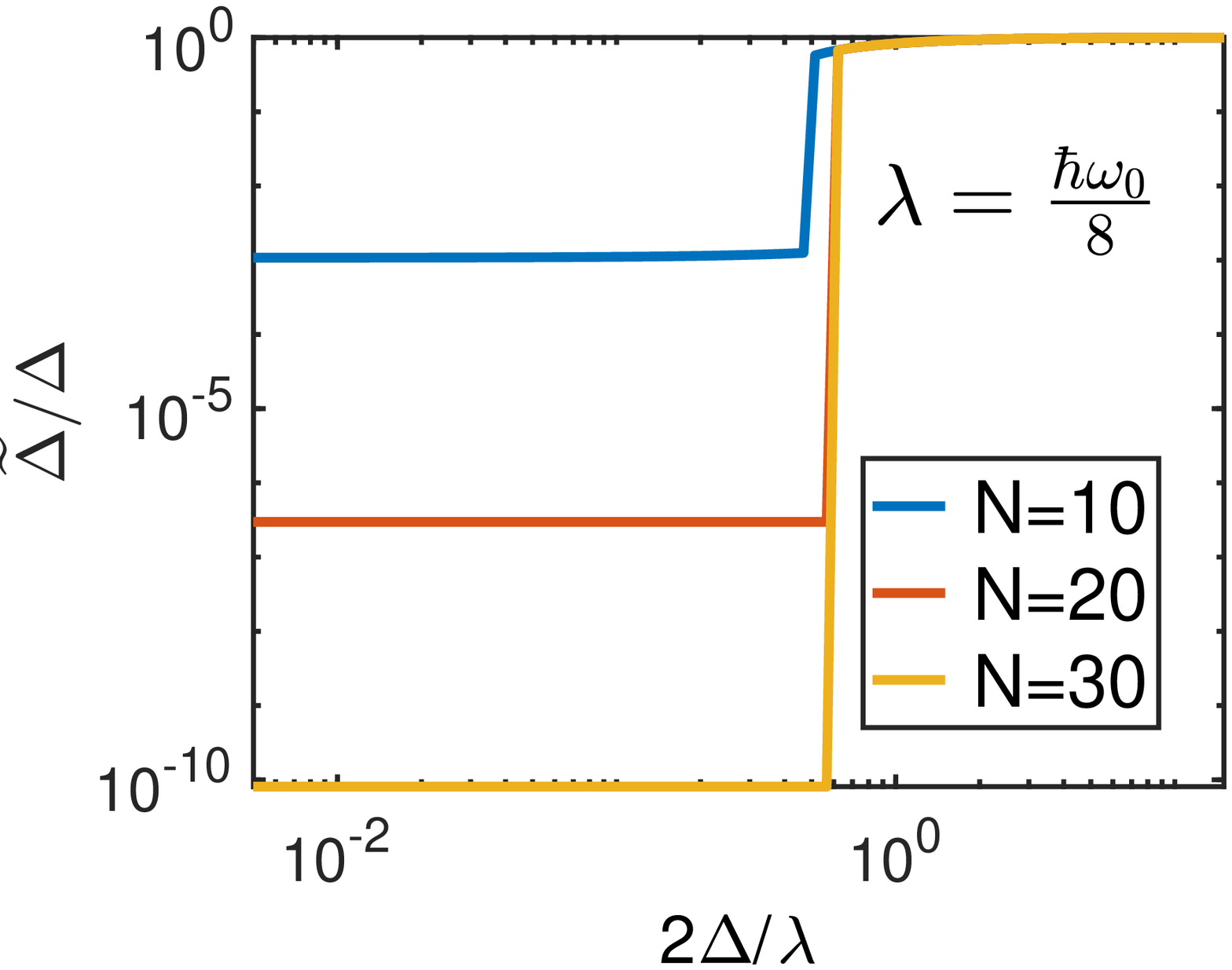}}
\subfloat[]{\includegraphics[scale=0.4]{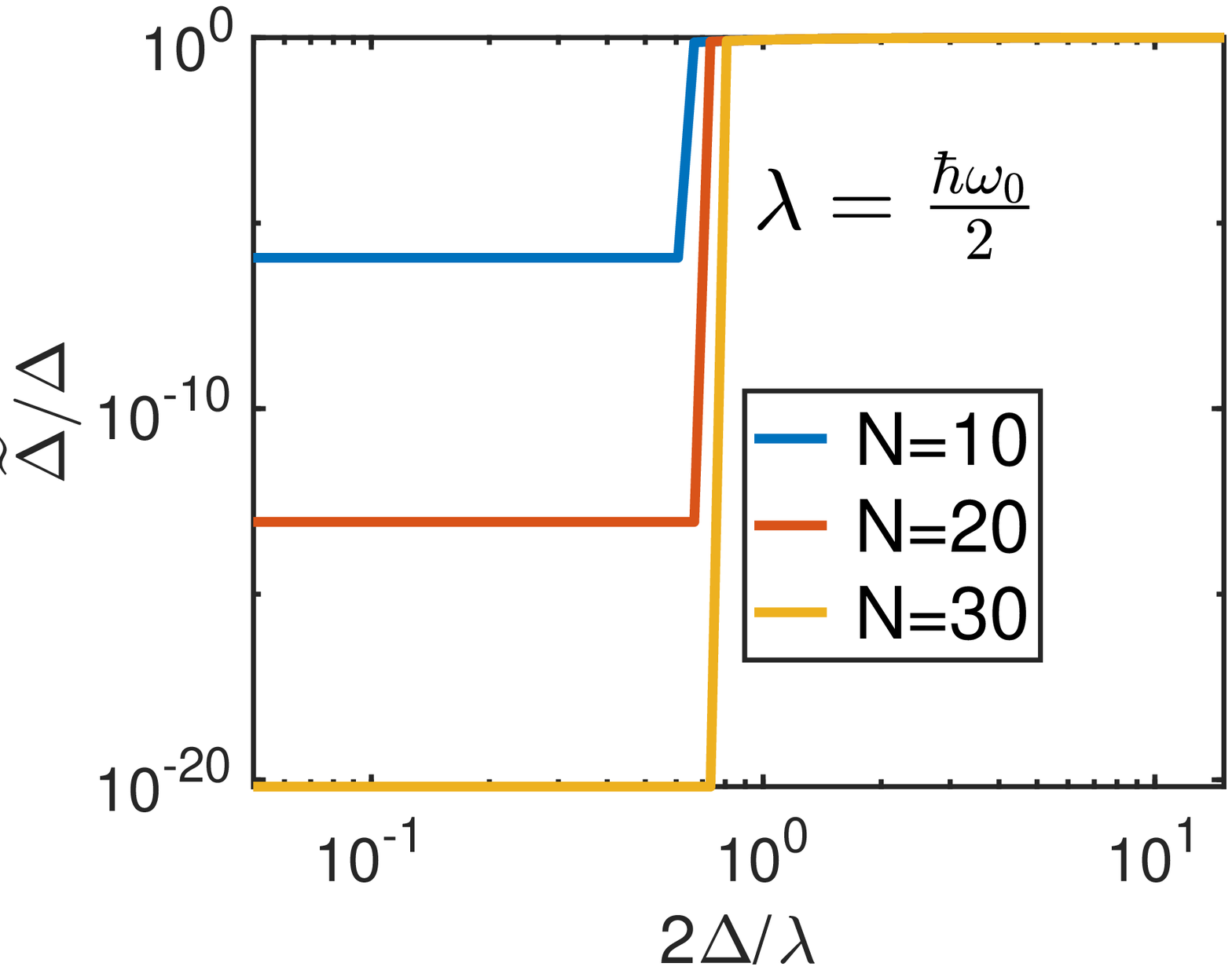}}
\\
\subfloat[]{\includegraphics[scale=0.4]{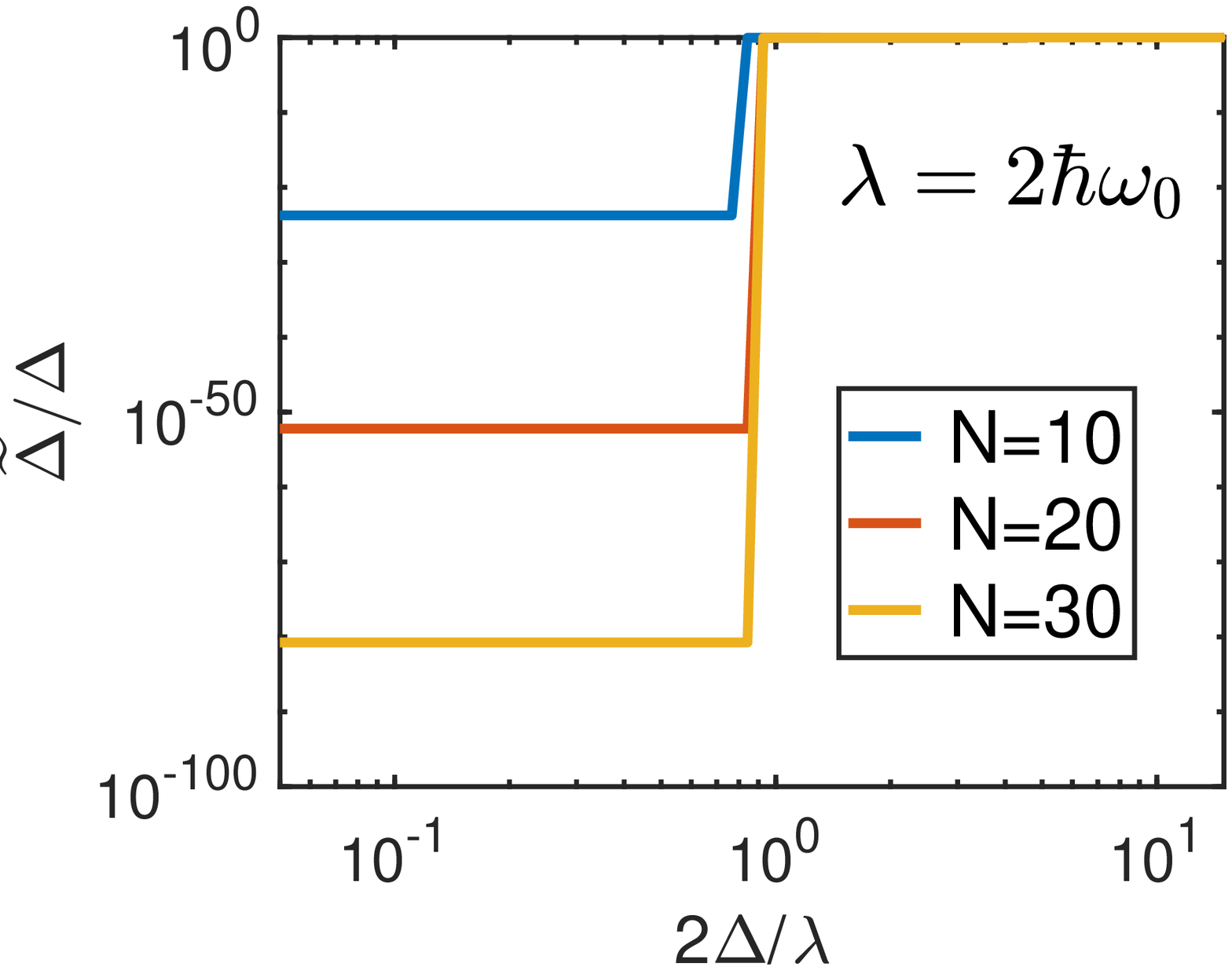}}
\subfloat[]{\includegraphics[scale=0.4]{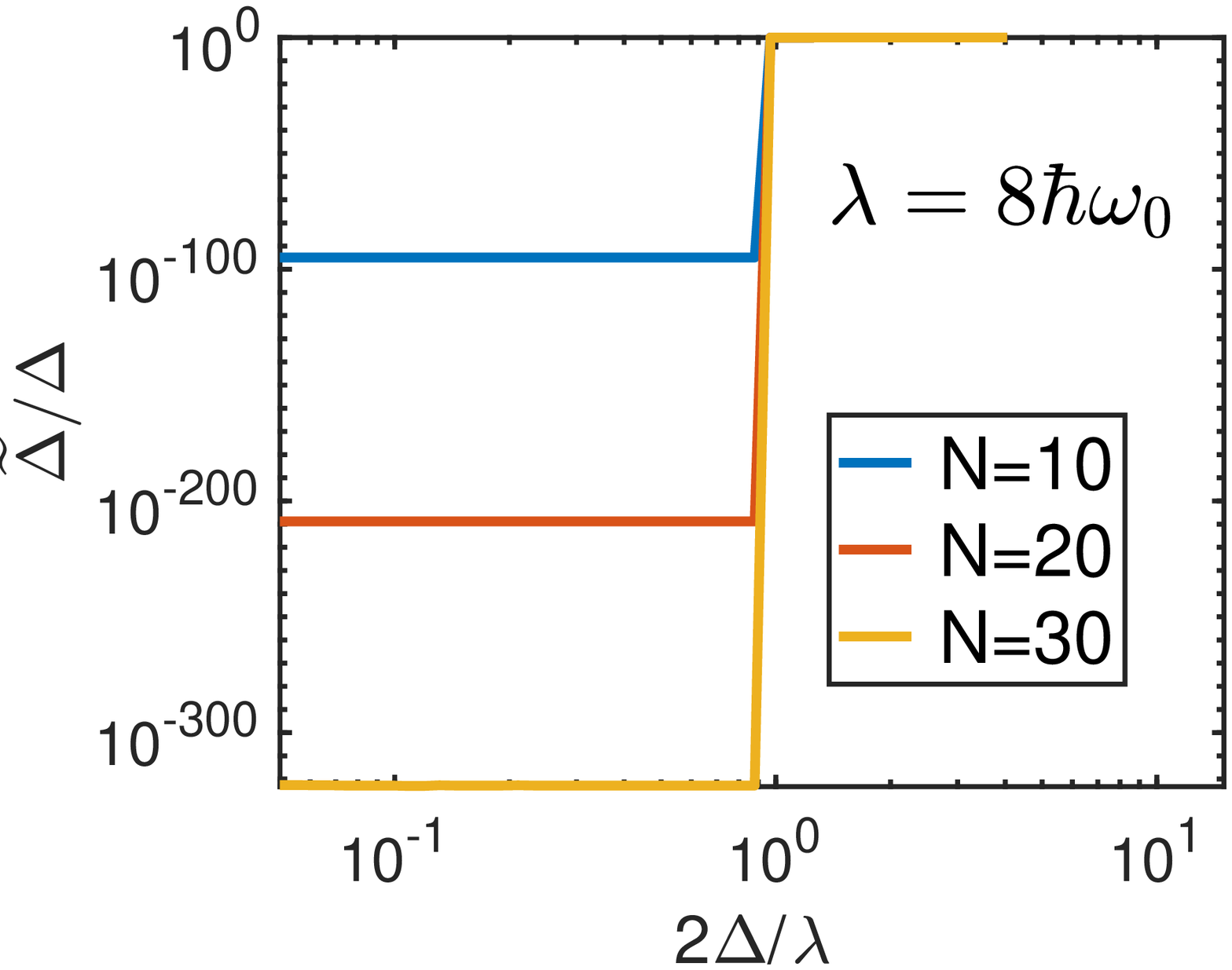}}
{\caption{\small  Relative renormalized hopping amplitude  $\tilde{\Delta}/\Delta$ vs. the relative bare hopping amplitude ($2\Delta/\lambda$) for different reorganization energies $\lambda/(\hbar\omega_{0})=1/8, 1/2, 2, 8$ shown in graphs {\bf a-d.} respectively, in the case of site-dependent interactions (color online, curve's height decreases with increasing $N$).} 
\label{fig:TunSub}}
\end{figure}

The more advanced theory \cite{Chin11PolarSC} leads to the factor of two larger estimate for $\Delta_{c}$, while the situation is still a controversy \cite{PVojtaSubOhm12ErrorsConsp}. The solution of Eq. (\ref{eq:SCPolaron})  leads to the transition consistent with Eq. (\ref{eq:Delta_c1}) as shown in Fig. \ref{fig:TunSup}. 

In the opposite limit of $\lambda \geq \hbar\omega_{0}$ the transition should take place at $\Delta_{c} \approx \lambda/2$ due to the energy minimum expectations as discussed in Sec. \ref{sec:pol-ident} for the superohmic bath. Similarly to that case the effective amplitude jumps down dramatically at $\Delta<\lambda/2$ in agreement with the solution of Eq. (\ref{eq:SCPolaron}) shown in Fig. \ref{fig:TunSub}. At finite system sizes the effective coupling is much weaker in subohmic then in superohmic case because of the infrared divergence of the polaron exponent in the former case. Asymptotic behaviors of the effective coupling strength $\tilde{\Delta}$ are summarized in Table \ref{tab:renorm-diff}.  We do not show the continuum infinite size limit since in that limit the renormalized coupling strength approaches zero, which is hard to show in the logarithmic plot.

\begin{table}
 \caption{Renormalization of hopping amplitude for different site interactions at zero temperature limit.} 
 \label{tab:renorm-diff}
%  \begin{tabular}{|p{3.5cm}|p{3.5cm}|p{3.5cm}|}
\centering
\begin{tabular}{|l|c|c|c|r|}
  \hline
 $\lambda$ vs. $\hbar\omega_{0}$ &\multicolumn{2}{|c|}{$\lambda<\hbar\omega_{0}$}&\multicolumn{2}{|c|}{$\hbar\omega_{0}<\lambda$}\\
  \hline
$\Delta$   &$\Delta<\frac{\lambda^2}{\hbar\omega_{0}}$ & $\frac{\lambda^2}{\hbar\omega_{0}}<\Delta$ & $\Delta<\lambda/2$ & $\lambda/2 < \Delta$  \\
   \hline
$\tilde{\Delta}$ &  $0$ & $\Delta$ & $0$ & $\Delta$  \\ 
   \hline  
  \end{tabular}
\end{table}

%We do not investigate the temperature dependence of the coherent coupling strength $\tilde{\Delta}$. This dependence is significant because it describes  the temperature  dependence of a particle group velocity. To characterize it one should examine a system dynamics that is beyond the scope of our analytical consideration. 

\section{Particle transfer between adjacent sites: transport and  decoherence }
\label{sec:Decoh1}

The consideration of effective hopping amplitude $\tilde{\Delta}$ in Sec. \ref{sec:decoh} is related to the  coherent transport at temperature approaching $0$. Here we examine the effect of a finite temperature  within the spin-boson problem corresponding to the particle hopping between two adjacent sites considering the interaction with transverse vibration for superohmic and subohmic baths for identical and different site interactions. In our consideration we use the Fermi-Golden rule approach developed in Ref. \cite{LeggettPolaronRevModPhys,KP86}  using the hopping interaction Eq. (\ref{eq:CanTr2}) as a perturbation. The Fermi-Golden rule is not always applicable and its applicability limits will be established during its evaluation. We also consider infinite chain limit assuming that temperature exceeds minimum quantization energies. 

Within this approximation the transition rate between two adjacent sites $p$ and $p+1$  can be expressed as \cite{LeggettPolaronRevModPhys}
\begin{eqnarray}
r=\frac{1}{\hbar^2}\int_{-\infty}^{\infty}dt \left<\widehat{V}_{h}(t)\widehat{V}_{h}(0)\right>,  
\label{eq:FGR0}
\end{eqnarray}
where time dependence and averaging are taken over  non-interacting phonons and two particle states at a finite temperature $T$. Using Eq. (\ref{eq:CanTr2}) one can represent the integrand in Eq. (\ref{eq:FGR0}) in terms of a correlation function  $C(t)=\Delta^2<e^{i\widehat{s}(t)}e^{-i\widehat{s}(0)}>$. Since the operator $\widehat{s}$ is linear in phonon creation and annihilation operators $\widehat{b}^{\dagger}$ and $\widehat{b}$ the function $C(t)$ can be evaluated exactly as  $C(t)=e^{-F(t)}$ with the exponent $F(t)$ defined as $F(t)=<\widehat{s}^2>-<\widehat{s}(t)\widehat{s}(0)>$. 

Using our definition of the exponent $\widehat{s}$ and shifting the integral in Eq. (\ref{eq:FGR0}) as $t \rightarrow t+i\hbar/(k_{B}T)$ \cite{BixonJortner} one can express the function $F(t)$ for the shifted time as 
\begin{eqnarray}
F(t)=F_{1}+F_{2}(t), ~ 
F_{1}=\frac{\lambda}{2N}\sum_{k}\frac{|1\mp e^{2\pi ik/N}|^2\tanh(\beta \hbar\omega_{k}/4)}{\hbar\omega_{k}},
\nonumber\\
F_{2}(t)=\frac{\lambda}{2N}\sum_{k}\frac{|1\mp e^{2\pi ik/N}|^2(1-\cos(\omega_{k} t))}{\hbar\omega_{k}\sinh(\beta\hbar\omega_{k}/2)}, ~ \beta=\frac{1}{k_{B}T}, 
\label{eq:FGRTr1}
\end{eqnarray}
where the minus sign corresponds to site independent and plus sign to site dependent interactions with transverse phonons. These expressions are subject to the frequency constraint limiting contributing frequencies by the inequality  $\omega_{k} \geq  r$ \cite{LeggettPolaronRevModPhys}  and the higher order corrections of the perturbation theory should be negligible. 

These conditions can be validated expressing the transition rate $r$ as 
\begin{eqnarray}
r \approx \frac{\Delta(T)^2}{\hbar k_{2}}, 
\label{eq:GenExpr}
\end{eqnarray} 
where $\Delta(T)=\Delta e^{-F_{1}/2}$ is the temperature dependent effective hopping amplitude and $k_{2}$ defines the decay rate of the correlation function due to time dependent part $e^{-F(t)}$. One can define this rate setting 
\begin{eqnarray}
F_{2}(1/k_{2})=1. 
\label{eq:DecR1}
\end{eqnarray}  
This rate represents the phase decoherence rate due to the interaction of particle with transverse phonons \cite{LeggettPolaronRevModPhys}. 

The Fermi Golden rule turns out to be applicable under the condition of large decoherence rate compared to the effective hopping and transport rates, which requires 
\begin{eqnarray}
\Delta(T) < k_{2}.
\label{eq:DecR1a}
\end{eqnarray}  
This condition  yields $r < k_{2}$ according to Eq. (\ref{eq:GenExpr}). If this condition is satisfied then the frequency constraint $\omega_{k} \geq  r$ can be neglected since the significant frequencies are of order of $k_{2}$. One can show that the higher order perturbation theory corrections are small in the factor  $\Delta(T)/k_{2}$ \cite{LeggettPolaronRevModPhys}.  Below we assume that Eq. (\ref{eq:DecR1}) is satisfied.

If Eq. (\ref{eq:DecR1}) is satisfied all calculations for functions determining decoherence and transport rate in Eq. (\ref{eq:FGRTr1})can be performed  ignoring lower frequency cutoffs. Since the calculations of decoherence and relaxation rates are straightforward and mostly performed earlier (see Ref. \cite{LeggettPolaronRevModPhys}) we give below the summary  of analytical, asymptotic behaviors for decoherence rate $k_{2}$ and temperature dependent effective hopping amplitude $\Delta(T)$. The transport rate can be evaluated using Eq. (\ref{eq:GenExpr}). 

First we give a summary of behaviors for the case of site-independent interactions. Three distinguishable temperature dependencies of decoherence rate are given below for low, intermediate and high temperatures in the form
\begin{eqnarray}
   k_{2}\approx 
  \begin{cases}
     \pi^5 \omega_{0} \left(\frac{\lambda}{\hbar\omega_{0}}\right)^2\left(\frac{k_{B}T}{\hbar\omega_{0}}\right)^2, ~ k_{2}< {\rm min} (\omega_{0}, k_{B}T/\hbar), ~ \text{\cite{LeggettPolaronRevModPhys}}, \\  \frac{\pi^2}{2\sqrt{6}}\omega_{0}
   \left(\frac{\lambda}{\hbar\omega_{0}}\right)^{1/2}\left(\frac{k_{B}T}{\hbar\omega_{0}}\right)^{3/2}, ~ \frac{k_{B}T}{\hbar}<k_{2}<\omega_{0}, ~ \& ~\hbar\omega_{0}< \lambda, \\
    \frac{\sqrt{k_{B}T\lambda}}{\hbar}, ~ \omega_{0} < k_{2}, ~ \text{Marcus's regime, \cite{Marcus2,BixonJortner} }.
  \end{cases}  
  \label{eq:Ulong} 
\label{eq:k2SummOdent}
\end{eqnarray}
The low temperature asymptotic behavior of decoherence rate will be used below in Sec. \ref{sec:exp} for the analysis of vibrational energy transport in poly-ethylene glycol oligomers \cite{ab20PegsExp}.

The remarkable renormalization of hopping amplitude takes place only for large reorganization energies, $\hbar\omega_{0} < \lambda$; otherwise one can set $\Delta(T) \approx \Delta$. For that case  using  Eq. (\ref{eq:FGRTr1}) we get 
\begin{eqnarray}
   \Delta(T)\approx 
  \begin{cases}
     \tilde{\Delta} , ~ k_{2}<\frac{k_{B}T}{\hbar},\\  \tilde{\Delta} \exp\left[7.5 
   \frac{\lambda}{\hbar\omega_{0}}\left(\frac{k_{B}T}{\hbar\omega_{0}}\right)^{1/2}\right], ~ \frac{k_{B}T}{\hbar}<k_{2}<\omega_{0}, ~ \& ~\hbar\omega_{0}< \lambda, \\
    \Delta e^{-\frac{\lambda}{4k_{B}T}}, ~ \omega_{0} < k_{2}, ~ \text{Marcus's regime, \cite{Marcus2,BixonJortner}  }.
  \end{cases}  
  \label{eq:UlongA} 
\label{eq:DeltaSummOdent}
\end{eqnarray}

In the case of subohmic bath, corresponding to site-dependent interactions the temperature dependencies of decoherence rate are weaker than in the superohmic case. They can be summarized as following
\begin{eqnarray}
   k_{2}\approx 
  \begin{cases}
    2.1 \omega_{0}  \left(\frac{\lambda}{\hbar\omega_{0}}\right)^{1/2}\left(\frac{k_{B}T}{\hbar\omega_{0}}\right)^{3/4}. ~ \frac{k_{B}T}{\hbar}< k_{2},  \\
  \left(\frac{2\sqrt{2\pi}}{3}\right)^{2/3} \omega_{0} \left(\frac{\lambda}{\hbar\omega_{0}}\right)^{2/3}\left(\frac{k_{B}T}{\hbar\omega_{0}}\right)^{2/3}, ~ \frac{k_{B}T}{\hbar}<k_{2}<\omega_{0} ~ \& ~\hbar\omega_{0}< \lambda, ~ \text{\cite{LeggettPolaronRevModPhys}} \\
    \frac{\sqrt{k_{B}T\lambda}}{\hbar}, ~ \omega_{0} < k_{2}, ~ \text{Marcus's regime, \cite{Marcus2,BixonJortner} }.
  \end{cases}  
\label{eq:k2SummDiff}
\end{eqnarray}

The renormalization of a hopping amplitude with the temperature is significant only for bare hopping amplitude smaller than the threshold, $\Delta<\Delta_{c}$. In that case the temperature dependence of hopping amplitude takes the form 
\begin{eqnarray}
   \Delta(T)\approx 
  \begin{cases}
    \Delta\exp\left(-\frac{\lambda}{\hbar\omega_{0}}\sqrt{\frac{\hbar\omega_{0}}{k_{B}T}}\right) , ~ \frac{k_{B}T}{\hbar}<\hbar\omega_{0},, ~  \text{\cite{LeggettPolaronRevModPhys}} \\  
    \Delta e^{-\frac{\lambda}{4k_{B}T}}, ~ \hbar\omega_{0} < k_{B}T, ~ \text{Marcus's regime, \cite{Marcus2,BixonJortner}  }.
  \end{cases}  
\label{eq:DeltaSummDiff}
\end{eqnarray}

%This estimate completes the general analysis 

\section{Decoherence of vibrational energy transport  in PEGs.}
\label{sec:exp}

% Choose acceptable modes, understand which is one is better
% Correct expression for reorganization energy, modify text
% Modify answers to the Referee
% Modfy conclusion. 
% Work on Sec. 4. 

The present theoretical investigation of a particle interacting with transverse phonons has been inspired by the measurements of temperature dependent transport of vibrational energy via polyethylene glycol oligomer chains (PEGs) \cite{ab20PegsExp}. The energy transport has been initiated by a single photoexcitation of azido group stretching mode, followed by  formation of a wavepacket (WP) propagating through  the chain. After reaching the opposite end of the molecule WP  decays by means of internal vibrational relaxation (IVR) shifting the frequency of the asymmetric C=O stretching mode of the succinimide ester end group, used as a reporter (see Fig. \ref{fig:PEGs}). The energy propagates through optical phonon bands and optical phonons can be considered as propagating particles interacting with transverse acoustic phonons.

Under the present experimental conditions each photoexcited molecule has around one optical phonon. Indeed, since the quantization energies for optical bands  studied experimentally \cite{ab15ballistictranspexp} exceed $1000$cm$^{-1}$ that is much larger than the thermal energy  these modes are populated only due to external excitations. Since a second quantum absorption is improbable as a higher order nonlinear process we have one optical phonon per each photoexcited molecule.  

%Since the quantization energy for the excited mode ($2200$cm$^{-1}$) and optical phonon bands ($\sim 1000$cm$^{-1}$  \cite{ab15JPCExpDec}) remarkably exceeds the thermal energy the optical phonon transport can be considered as a single particle transport affected by the interaction with acoustic phonon bath so the present theory should be applicable there. 

 It was found experimentally that the transport velocity decreases with increasing  temperature. The experimental data  have been analyzed using the density matrix formalism applied to these optical phonons featuring the decoherence rate obeying the law $k_{2}=1/T_{2} \propto T^2$.   Reasonable quantitative interpretation of experiment has been obtained choosing $k_{2}$ as
\begin{eqnarray}
k_{2}=0.53\cdot \left(\frac{T}{290}\right)^2 ps.^{-1}. 
\label{eq:ExpRes}
\end{eqnarray}

We wish to compare this result with the predictions of the present work  given by Eq. (\ref{eq:k2SummOdent}). To determine the parameters in Eq. (\ref{eq:k2SummOdent}) we used third order anharmonic force constants $A_{\alpha\beta\gamma}$ evaluated using  DFT
(B3LYP/6-311G++(d,p)) geometry optimization by the computational chemistry software package  Gaussian \cite{Gaussian} for PEG oligomers studied  experimentally  \cite{ab20PegsExp,PandeyLeitner17ThermPEGs,ab16jpcPEGs}. 

Let us approximate the PEG molecule by the $N$ site chain. According to Eq. (\ref{eq:k2SummOdent}) the decoherence rate is determined by the characteristic maximum frequency $\omega_{0}$  of transverse (bending)  vibrational modes and the local reorganization energy $\lambda$. One can estimate the maximum frequency using the frequency $\omega_{1}$ of the lowest energy bending mode that is connected to the frequency $\omega_{0}$ as $\omega_{0} \approx N^2\omega_{1}$. This frequency is related to the Gaussian output frequency $\nu_{1}$ expressed in inverse cm as \cite{Barone}
\begin{eqnarray}
\omega_{0}=2\pi c N^2 \nu_{1},
\label{eq:FrBendMode}
\end{eqnarray}
where $c$ is the speed of light expressed in cm per sec. 

Optical phonons occupying the delocalized normal mode  $\alpha$ are coupled to the bending modes by means of the anharmonic interaction. The interaction responsible for reorganization and decoherence is defined by the third order anharmonic force constants $A_{\alpha\alpha 1}$ that are also expressed in cm$^{-1}$  within the Gaussian output. The reorganization energy  $\lambda_{\alpha}$ can be defined similarly to  Eq. (\ref{eq:part-phint}) as 
\begin{eqnarray}
 \frac{2\pi c\hbar A_{\alpha \alpha 1}}{\sqrt{8}}=\frac{\sqrt{\lambda_{\alpha}\hbar\omega_{0}}}{\sqrt{2}N^{3/2}}.   
\label{eq:AnhConnect}
\end{eqnarray} 
Consequently,  the normal mode $a$ reorganization energy can be expressed using Eq. (\ref{eq:sp-ph}) as 
\begin{eqnarray}
\lambda_{\alpha}=N\pi c\hbar \frac{A_{nn1}^2}{\nu_{1}}. 
\label{eq:reorgexpComp}
\end{eqnarray}

The mode $a$ is delocalized over $N$ sites. The local reorganization energy $\lambda$ used in Eq. (\ref{eq:sp-ph}) should exceed the energy $\lambda_{\alpha}$ by a factor of $N$ because of the standard gain in reorganization due to localization \cite{ab08jcpDNALocaliz}  so we get 
\begin{eqnarray}
\lambda=\frac{N^2\pi c\hbar}{2}\frac{A_{nn1}^2}{\nu_{1}}. 
\label{eq:reorgexpCompA}
\end{eqnarray}

Using Eq. (\ref{eq:k2SummOdent}) one can approximately express the decoherence rate in terms of the Gaussian output parameters as 
\begin{eqnarray}
k_{2} \approx \frac{\pi^5}{16}\frac{2\pi c\nu_{1}}{N^2}
\left(\frac{k_{B}T}{2\pi \hbar c\nu_{1}}\right)^2
\left(\frac{A_{\alpha\alpha 1}}{\nu_{1}}\right)^4. 
\label{eq:DecohExp}
\end{eqnarray}

To compare the prediction, Eq. (\ref{eq:DecohExp}), with the experimentally derived behavior, Eq. (\ref{eq:ExpRes}), we considered PEG oligomers with $N=1, 2$ and $3$ monomers (see Fig. \ref{fig:PEGs}). The maximum size is restricted by the long computational time and poor convergence; there are couple of negative non-linear frequencies already for $N=3$ that is hard to avoid using standard geometry optimization. Since the frequencies of the modes of interest are positive we believe that the consideration still makes sense. Oligomers with $N>3$ require special consideration using tight geometry optimization to avoid negative frequency \cite{Gaussian}, which is postponed for future work.

The lowest frequency transverse (bending) modes have been identified visually using the program Molden \cite{Molden1,Molden2}. The lowest frequency bending mode is the fourth mode out of $54$ with  the frequency $\nu_{1}=56$ cm$^{-1}$  for $N=1$,  the third mode out of $75$ with  the frequency $\nu_{1}=33$ cm$^{-1}$ for $N=2$ and the third mode out of $96$ with  the frequency $\nu_{1}=22$ cm$^{-1}$  for $N=3$. Mode frequencies do not follow $N^{-2}$ law. This can be the consequence of level repulsion with torsional modes located at lower frequencies. Assuming that this effect is less significant for $N=3$ one can estimate the maximum frequency $\omega_{0}$ to be around $200$ cm$^{-1}$. Unfortunately, the calculations for larger molecules where bending modes will have have lower frequencies than other modes are problematic.

\begin{figure}
\includegraphics[scale=0.3]{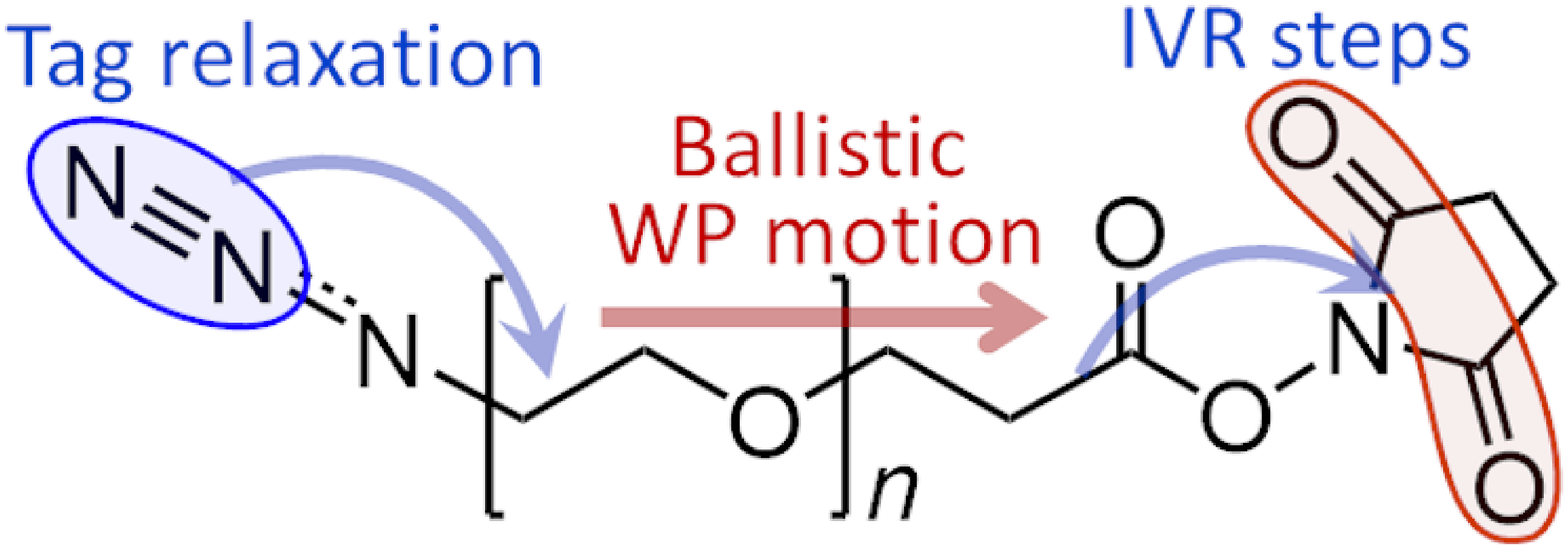} 
\caption{Energy transport in PEG oligomers investigated in Ref. \cite{ab20PegsExp}.}
\label{fig:PEGs}
\end{figure}
 
 The interaction of bending mode with several delocalized optical phonon modes  is sufficiently strong ($1$cm$^{-1} \leq A_{\alpha\alpha 1}<10$cm$^{-1}$)   to  have the decoherence rate (see Eq. (\ref{eq:DecohExp}))) comparable to the experiment based estimate of Eq. (\ref{eq:ExpRes}). The corresponding reorganization energy can be around $1-20$ cm$^{-1}$.  

 The best match for decoherence rate  is found for  modes with  frequencies around $1000$cm$^{-1}$ that correspond to delocalized $C-C$  stretching vibrations and with frequencies around $1300$cm$^{-1}$ for collective hydrogen vibrations (wagging). The anharmonic interactions, $A$, and frequencies, $\nu$, for these modes are given by for N=1: $A=-0.94$cm$^{-1}$ , $\nu=953$cm$^{-1}$ (stretching)  and $A=0.99$cm$^{-1}$, $\nu=1308$cm$^{-1}$ (wagging), for N=2: $A=-1$cm$^{-1}$ , $\nu=1011$cm$^{-1}$ and $A=0.62$cm$^{-1}$ , $\nu=1039$cm$^{-1}$ (stretching),  and $A=-0.79$cm$^{-1}$, $\nu=1351$cm$^{-1}$ and $A=0.79$cm$^{-1}$, $\nu=1358$cm$^{-1}$   (wagging), for N=3: $A=4.08$cm$^{-1}$ , $\nu=1020$cm$^{-1}$ (stretching),  and $A=1.76$cm$^{-1}$, $\nu=1302$cm$^{-1}$, and  $A=2.38$cm$^{-1}$, $\nu=1305$cm$^{-1}$(wagging). 
We consider wagging optical phonons as the most suitable candidates for energy transport since they possess a narrow band with the width comparable to the one ($4\Delta \sim 2\pi \hbar c \cdot 20$cm$^{-1}$) estimated experimentally  
  \cite{ab14PerFluoroAlkExp,ab20PegsExp}. 

The increase in the number of coupled modes is a natural trend that is seen for $N=1$ and $N=2$ while the results for $N=3$ can be not sufficiently accurate because of the poor convergency. The decoherence rate can be underestimated for longer sequences  because for the ballistic transport the decoherence is significant not only for nearest neighbors but for more separated sites where it can increase as a squared distance between them  (see Eq. (\ref{eq:FGRTr1})). 
 
One should notice that according to Ref. \cite{ab20PegsExp} the decoherence rate is smaller than the hopping amplitude $\Delta/\hbar \sim 1$ps$^{-1}$ ($5$cm$^{-1}$) so formally the restriction in Eq. (\ref{eq:DecR1}) is violated. However, since the bandwidth is smaller than the thermal energy $k_{B}T$ \cite{ab20PegsExp} for all probed temperature except for the lowest one  the estimate can still make sense as input into the equation for the density matrix used in Ref. \cite{ab20PegsExp}. Also since two rates are of the same order of magnitude we believe that the estimate in Eq. (\ref{eq:k2SummOdent}) should be valid at least qualitatively. The other constraints  formulated in Eq. (\ref{eq:k2SummOdent}) are clearly satisfied for all experimental data since the maximum decoherence rate   is around $1$ps$^{-1}$ while the maximum bending mode frequency $\omega_{0}$ and associated thermal frequency $k_{B}T/\hbar$ are at least $10$ times larger.

%since the extracted decoherence rate is smaller than both characteristic frequency $\omega_{0}$ since it is less than the minimum bending mode frequency ($2\pi c \nu_{1} $) and ``thermal" frequency  $k_{B}T/\hbar$. Also the reorganization energies $\lambda \sim 10$cm$^{-1}$  are smaller than the quantization energy $\hbar\omega_{0} \sim 100$cm$^{-1}$ that permits us to neglect the renormalization of the hopping amplitude. That supports the expected $T^{2}$ dependence of $k_{2}$. 

 %The other constraint of Eq. (\ref{eq:narrBand}) for the minimum temperature is well satisfied for all experimental temperatures except the lowest one.  

It is harder to identify other transverse acoustic modes and estimate the optical phonon interactions with them because the molecules under consideration  are short and these modes can be mixed up with other modes, while the experiment is made for longer molecules where the calculation of anharmonic interactions is problematic because of the poor convergence, We still hope that our calculations can be used for a crude estimate of а decoherence rate that is the main target of the present section.

% Gaussian data
% express all in inverse cm

\section{Conclusions}

% Compare two regimes of renormalization. 

The present work demonstrates significance of transverse vibrations for a particle propagating in one-dimensional systems. It turns out that the polaron effect and decoherence of particle transport can be stronger due to its interaction with transverse phonons compared to that of longitudinal phonons. Particularly we found  that in case of site dependent vibrational interactions transverse phonons act like a subohmic bath, while for identical interactions they act like as a superohmic bath. Two dimensional transverse vibrations are expected to serve as an ohmic bath.

% Coherence, quantum phase transition
The site dependent interaction with transverse phonons leads to the whole branch of phenomena known for subohmic baths including the quantum phase transition accompanied by the vanishing of transport at zero temperature and fast decoherence  at low temperature. The sharp quantum phase transition can be of interest for molecular devices, including molecular switches. 
In the superohmic case of site independent interactions the   decoherence is characterized by the rate decreasing with the temperature as $T^{2}$. 

The results for site-independent interactions are applied to recently investigated energy transport through the polyethylene glycols  \cite{ab20PegsExp}. The decoherence rate extracted from these measurements is consistent with the theoretical estimates of a present work for certain modes. Future experimental transport studies in  longer chains are desirable to provide further verification for the present theory. %Particularly diffusion coefficient at low temperature should be inversely proportional to the decoherence rate \cite{ab15jcp} leading to $D \propto T^{-2}$ temperature dependence that can be probed in the pure diffusion transport regime. 

% Decoherence rate 
% comparison to the experiment limited accuracy, small variation in wavevector q_1 modifies things by orders of magnitude; needs for future studies

\begin{acknowledgments}
This work is supported by the National Science Foundation (CHE-1900568).

We acknowledge Nikolaj Prokofev and Igor Rubtsov for useful discussions, and Robert Mackin and Layla Qasim for help in calculation of anharmonic interactions.

\end{acknowledgments}

\section{Data Availability}

The data that support the findings of this study are available from the corresponding author upon reasonable request.
\bibliography{Vibr}
\end{document}